\documentstyle[12pt,epsfig]{article}
\makeatletter
\def\slash#1{{\mathpalette\c@ncel{#1}}} 
\makeatother

\newcommand\beq{\begin{eqnarray}}
\newcommand\eeq{\end{eqnarray}}

\newcommand\la{\langle}
\newcommand\ra{\rangle}

\begin{document}
\begin{flushright}
\end{flushright}
\vspace*{15mm}
\begin{center}
{\Large \bf Three-gluon contribution to the single spin asymmetry\\[7pt]
for light hadron production in pp collision}
\vspace{1.5cm}\\
 {\sc Hiroo Beppu$^1$, Koichi Kanazawa$^{1,2}$, Yuji Koike$^3$ and Shinsuke Yoshida$^{4,5}$}
\\[0.7cm]
\vspace*{0.1cm}{\it $^1$ Graduate School of Science and Technology, Niigata University,
Ikarashi, Niigata 950-2181, Japan}\\

\vspace{0.2cm}

{\it $^2$ Department of Physics, Barton Hall, Temple University, Philadelphia, Pennsylvania 19122, USA}

\vspace{0.2cm}

{\it $^3$ Department of Physics, Niigata University, Ikarashi, Niigata 950-2181, Japan}

\vspace{0.2cm}
{\it $^4$ Theoretical Research Division, Nishina Center, RIKEN, Wako 351-0198, Japan}

\vspace{0.2cm}
{\it $^5$ Physics Department, Brookhaven National Laboratory, Upton, NY 11973, USA}
\\[3cm]

{\large \bf Abstract} \end{center}

We study the twist-3 three-gluon contribution to the single spin asymmetry in the
light-hadron production in pp collision in the framework of the collinear factorization.
We derive the corresponding cross section formula in the leading order with respect to
the QCD coupling constant.  
We also present a numerical calculation of
the asymmetry at the RHIC energy, using a model for the three-gluon correlation
functions suggested by the asymmetry 
observed in the $D$-meson production at RHIC.

\newpage
\section{Introduction}

The large single spin asymmetry (SSA) $A_N\equiv 
(\sigma^\uparrow -\sigma^\downarrow)/ (\sigma^\uparrow + \sigma^\downarrow)$ for
the light-hadron productions, $p^\uparrow p\to hX$ ($h=\pi,K,\eta$),
observed by the Relativistic Heavy Ion collider
(RHIC) at the Brookhaven National Laboratory
(BNL)\,\cite{Star2004}-\cite{Phenix2013} is the cleanest data
which can be analyzed by means of the twist-3 mechanism
in the framework of the collinear factorization.  
This process receives, in principle, four types of the
twist-3 contributions
depending on the type of the twist-3 correlation function
participating in the cross section; i.e.,
(i) twist-3 quark-gluon correlation functions in the polarized
nucleon\,\cite{QS1992}-\cite{KoikeTomita2009}, (ii) twist-3 three-gluon 
correlation functions in the 
polarized nucleon\,\cite{BJLO01}-\cite{KoikeYoshida2012}, (iii) twist-3 fragmentation 
function for the final 
hadron\,\cite{Ji1994}-\cite{MetzPitonyak2013}
and (iv) twist-3 quark-gluon correlation function in the initial unpolarized
nucleon\,\cite{KanazawaKoike2000}.
The last contribution (iv) was shown to be negligible due to the small
partonic hard cross sections\,\cite{KanazawaKoike2000E}. 
So far the data of $A_N$ for the light-hadron production has been analyzed
by assuming that the contribution (i) saturates the whole 
asymmetry\,\cite{KouvarisQiuVogelsangYuan2006,KanazawaKoike2010,KanazawaKoike2011,Kanazawa:2012kt}.
Although the gross feature of the observed asymmetry has been 
well reproduced by these analyses, 
the above (ii) and (iii)
could also become an important sources of the asymmetry.  
Furthermore, 
the sign of the quark-gluon correlation functions
in the above analysis is opposite to what is expected from the analysis of SSAs
observed in semi-inclusive deep inelastic scattering (SIDIS)\,\cite{Kang:2011hk}. 
Therefore it is important to perform a global analysis of the SSAs for variety of
processes including the whole contributions.  
The cross section formula for (iii) was derived only recently\,\cite{MetzPitonyak2013},
while the formula for (ii) is not available in the literature.

In our recent papers, 
we have derived the contribution from the twist-3 three-gluon
correlation functions to SSA for the $D$-meson productions in 
SIDIS, 
$ep^\uparrow \to eDX$\,\cite{BeppuKoikeTanakaYoshida2010,BeppuKoikeTanakaYoshida2012}, 
and the pp collision, $p^\uparrow p\to DX$\,\cite{KoikeYoshida2011}, 
and the Drell-Yan/direct-photon processes, 
$p^\uparrow p \to \gamma^{(*)} X$\,\cite{KoikeYoshida2012}.  
Using some models for the three-gluon correlation functions, 
we also studied its impact on the corresponding asymmetries  
at the energy of RHIC 
and the Electron-Ion-Collider (EIC),
and showed the sensitivity of the asymmetries to the form of the three-gluon
correlation functions.

The purpose of this paper is to derive the three-gluon contribution
to SSA for the light-hadron production
in the pp collision by applying the formalism developed for the above processes.
This will complete the corresponding leading order twist-3 cross section
together with the known result for the above contributions (i) and (iii).  
Our presentation in this paper will be brief, referring the readers
to \cite{BeppuKoikeTanakaYoshida2010,KoikeTanakaYoshida2011,KoikeYoshida2011,KoikeYoshida2012} 
for the detail of the calculation.  
After introducing the three-gluon correlation functions in section 2,
we present the twist-3 cross section in section 3.  In section 4,
we study its impact on the SSA in the light-hadron productions at the RHIC 
energy by using the model
in \cite{KoikeYoshida2011,KoikeYoshida2012}.  We will see the three-gluon correlation
disturbs the asymmetry reproduced by the quark-gluon correlation functions
in \cite{KanazawaKoike2010,KanazawaKoike2011}, which shows that
this process may be used to constrain the magnitude and the form
of the three-gluon correlations.    

\section{Three-gluon correlation functions in the transversely polarized nucleon}

As clarified in \cite{BJLO01,Braun09,BeppuKoikeTanakaYoshida2010},
there are two independent three-gluon correlation functions
in the transversely polarized nucleon, 
$O(x_1,x_2)$ and $N(x_1,x_2)$, 
which are the Lorentz-scalar functions of the longitudinal momentum fractions $x_1$ and $x_2$, 
defined as
\beq
&&\hspace{-0.8cm}O^{\alpha\beta\gamma}(x_1,x_2)
=-g(i)^3\int{d\lambda\over 2\pi}\int{d\mu\over 2\pi}e^{i\lambda x_1}
e^{i\mu(x_2-x_1)}\la pS_\perp
|d_{bca}F_b^{\beta n}(0)F_c^{\gamma n}(\mu n)F_a^{\alpha n}(\lambda n)
|pS_\perp\ra \nonumber\\
&&=2iM_N\left[
O(x_1,x_2)g^{\alpha\beta}\epsilon^{\gamma pnS_\perp}
+O(x_2,x_2-x_1)g^{\beta\gamma}\epsilon^{\alpha pnS_\perp}
+O(x_1,x_1-x_2)g^{\gamma\alpha}\epsilon^{\beta pnS_\perp}\right]
\label{3gluonO},\\
&&\hspace{-0.8cm}N^{\alpha\beta\gamma}(x_1,x_2)
=-g(i)^3\int{d\lambda\over 2\pi}\int{d\mu\over 2\pi}e^{i\lambda x_1}
e^{i\mu(x_2-x_1)}\la pS_\perp
|if_{bca}F_b^{\beta n}(0)F_c^{\gamma n}(\mu n)F_a^{\alpha n}(\lambda n)
|pS_\perp\ra \nonumber\\
&&=2iM_N\left[
N(x_1,x_2)g^{\alpha\beta}\epsilon^{\gamma pnS_\perp}
-N(x_2,x_2-x_1)g^{\beta\gamma}\epsilon^{\alpha pnS_\perp}
-N(x_1,x_1-x_2)g^{\gamma\alpha}\epsilon^{\beta pnS_\perp}\right], 
\label{3gluonN}
\eeq
where $F_a^{\alpha\beta}\equiv\partial^\alpha A^\beta_a
-\partial^\beta A^\alpha_a +gf_{abc}A_b^\alpha A_c^\beta$ is the gluon's
field strength, and we used the notation $F_a^{\alpha n}\equiv F_a^{\alpha \beta}n_{\beta}$
and $\epsilon^{\alpha pnS_\perp}\equiv \epsilon^{\alpha\mu\nu\lambda}p_\mu n_\nu 
S_{\perp\lambda}$
with the convention $\epsilon_{0123}=1$.  
$d^{bca}$ and $f^{bca}$ are the symmetric
and anti-symmetric structure constants of the color SU(3) group,
and we have suppressed the gauge-link operators which ensure the gauge invariance.
$p$ is the nucleon momentum, and
$S_\perp$ is the transverse spin vector of the
nucleon normalized as $S_\perp^2=-1$.
In the twist-3 accuracy, $p$ can be regarded as lightlike ($p^2=0$), 
and $n$ is another lightlike vector satisfying $p\cdot n=1$.  To be specific, 
we set $p^\mu=(p^+,0,\mathbf{0}_\perp)$, $n^\mu=(0,n^-, \mathbf{0}_\perp)$, and  
$S^\mu_\perp =(0,0, \mathbf{S}_\perp)$.
The nucleon mass $M_N$ is introduced to define 
$O(x_1,x_2)$ and $N(x_1,x_2)$ dimensionless.  The
decomposition (\ref{3gluonO}) and (\ref{3gluonN})
takes into account all the constraints from hermiticity, 
invariance 
under the parity- and time-reversal transformations and the permutation 
symmetry among the participating three gluon-fields.  The functions
$O(x_1,x_2)$ and $N(x_1,x_2)$ are real and have the following symmetry
properties,
\beq
&&O(x_1,x_2)=O(x_2,x_1),\qquad O(x_1,x_2)=O(-x_1,-x_2),\label{symO}\\
&&N(x_1,x_2)=N(x_2,x_1),\qquad N(x_1,x_2)=-N(-x_1,-x_2).\label{symN}  
\eeq
The functions $N(x_1,x_2)$ and $O(x_1,x_2)$ are, respectively, even and odd
under charge conjugation.

\section{Twist-3 cross section for $p^\uparrow p \to h X$ 
induced by the three-gluon correlation functions}

\begin{figure}[ht]
\begin{center}
  \includegraphics[height=4cm,width=5cm]{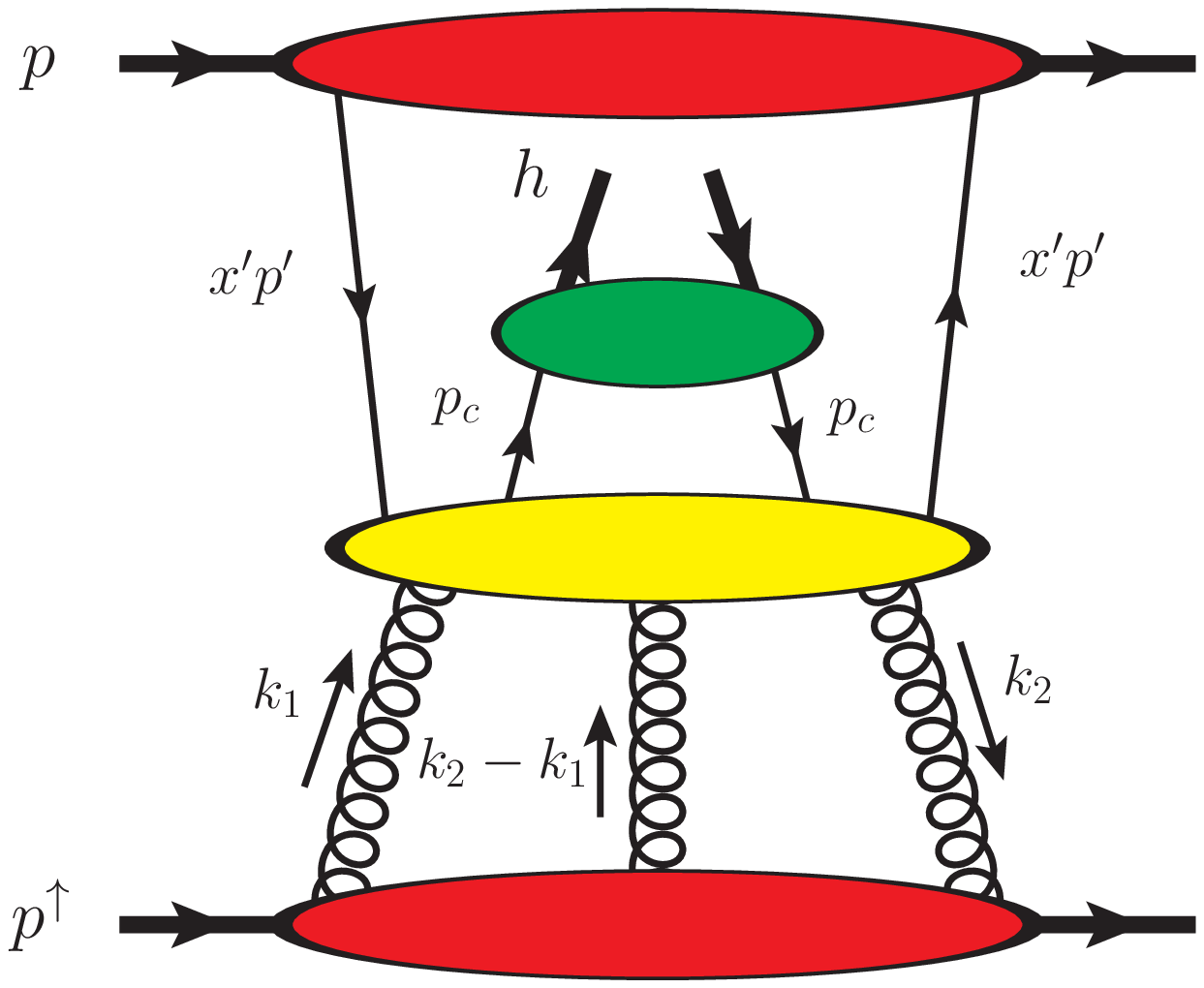}\hspace{1cm}
  \includegraphics[height=4cm,width=5cm]{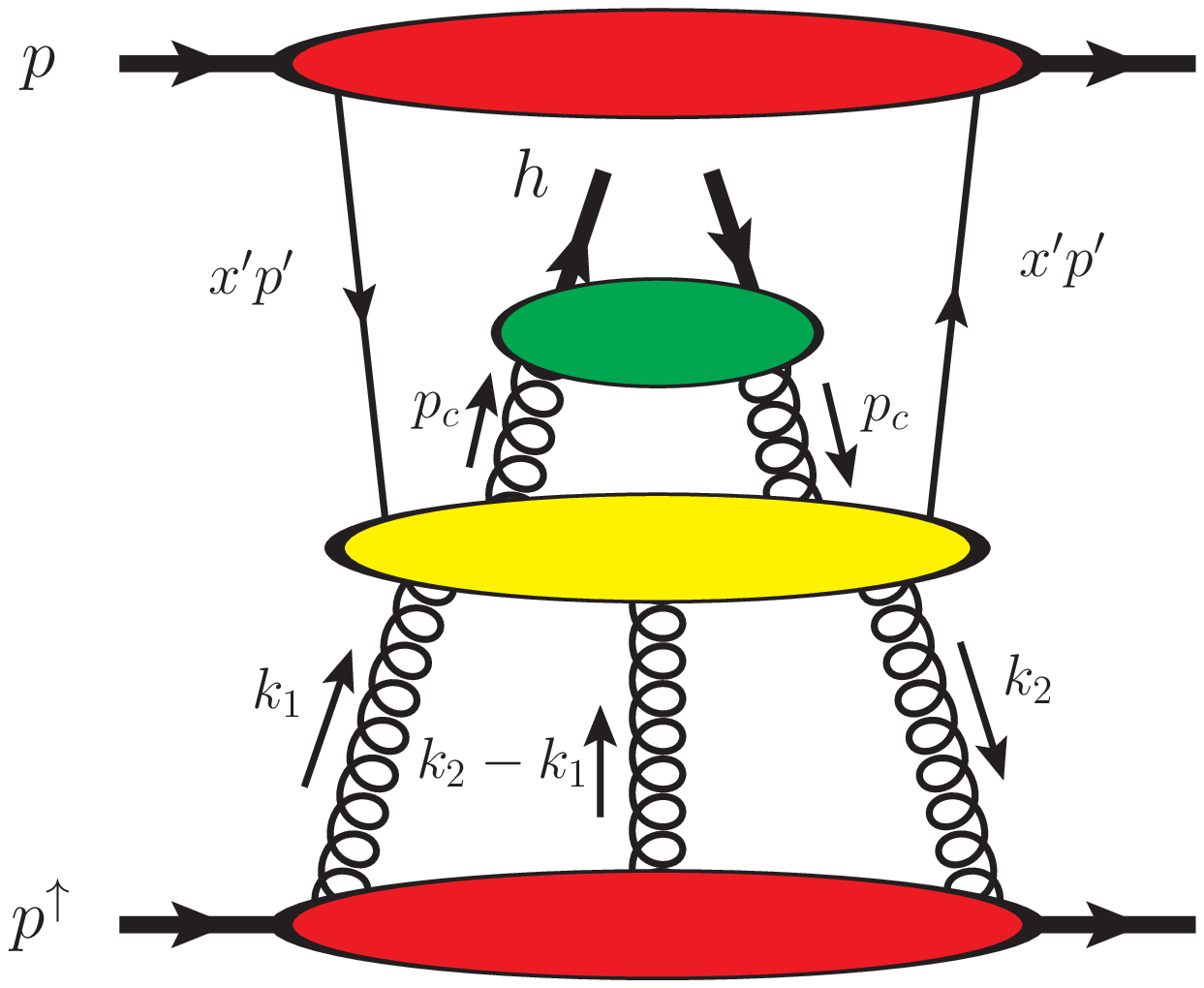}\\
  \vspace{1cm}
  \includegraphics[height=4cm,width=5cm]{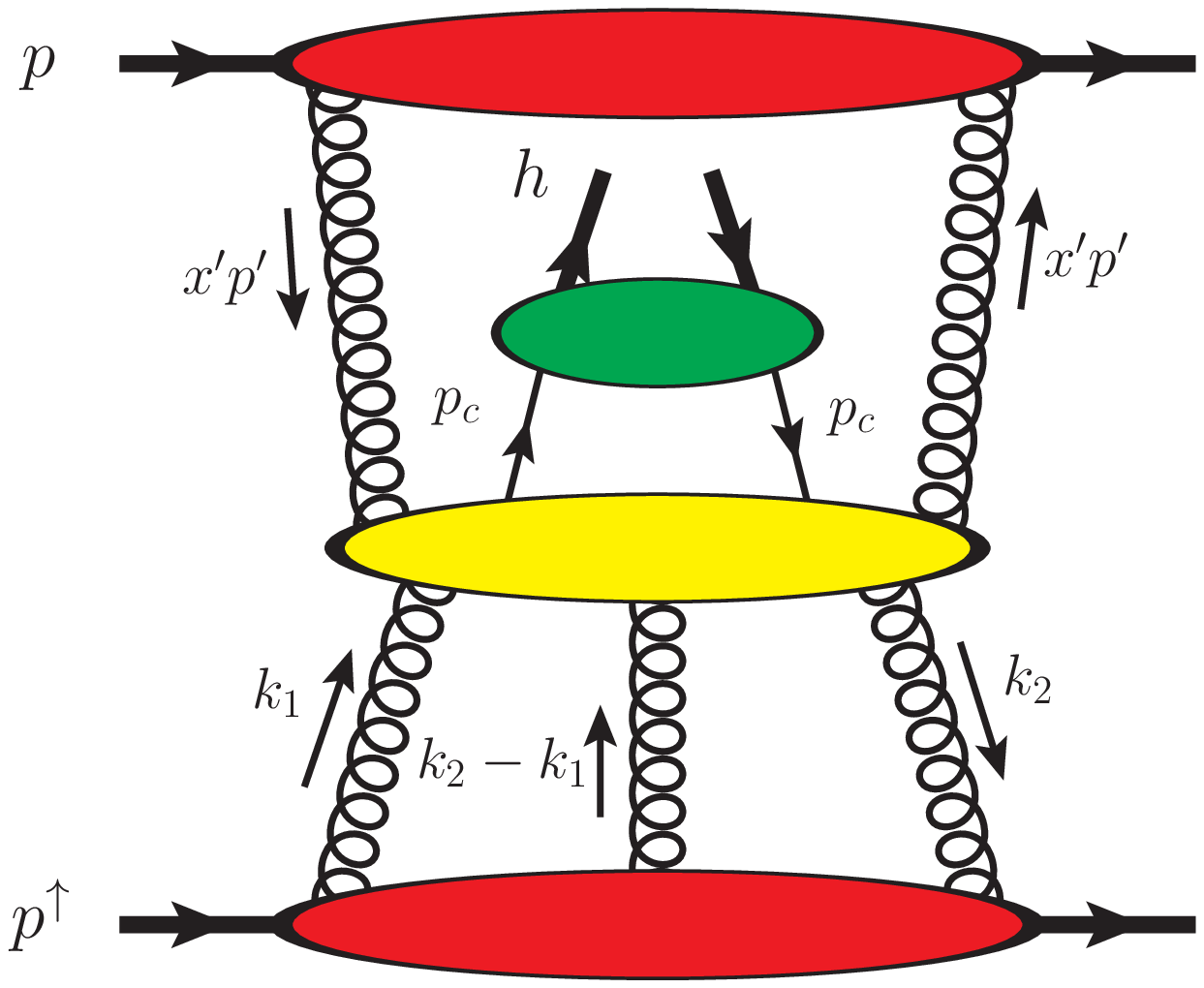}\hspace{1cm}
  \includegraphics[height=4cm,width=5cm]{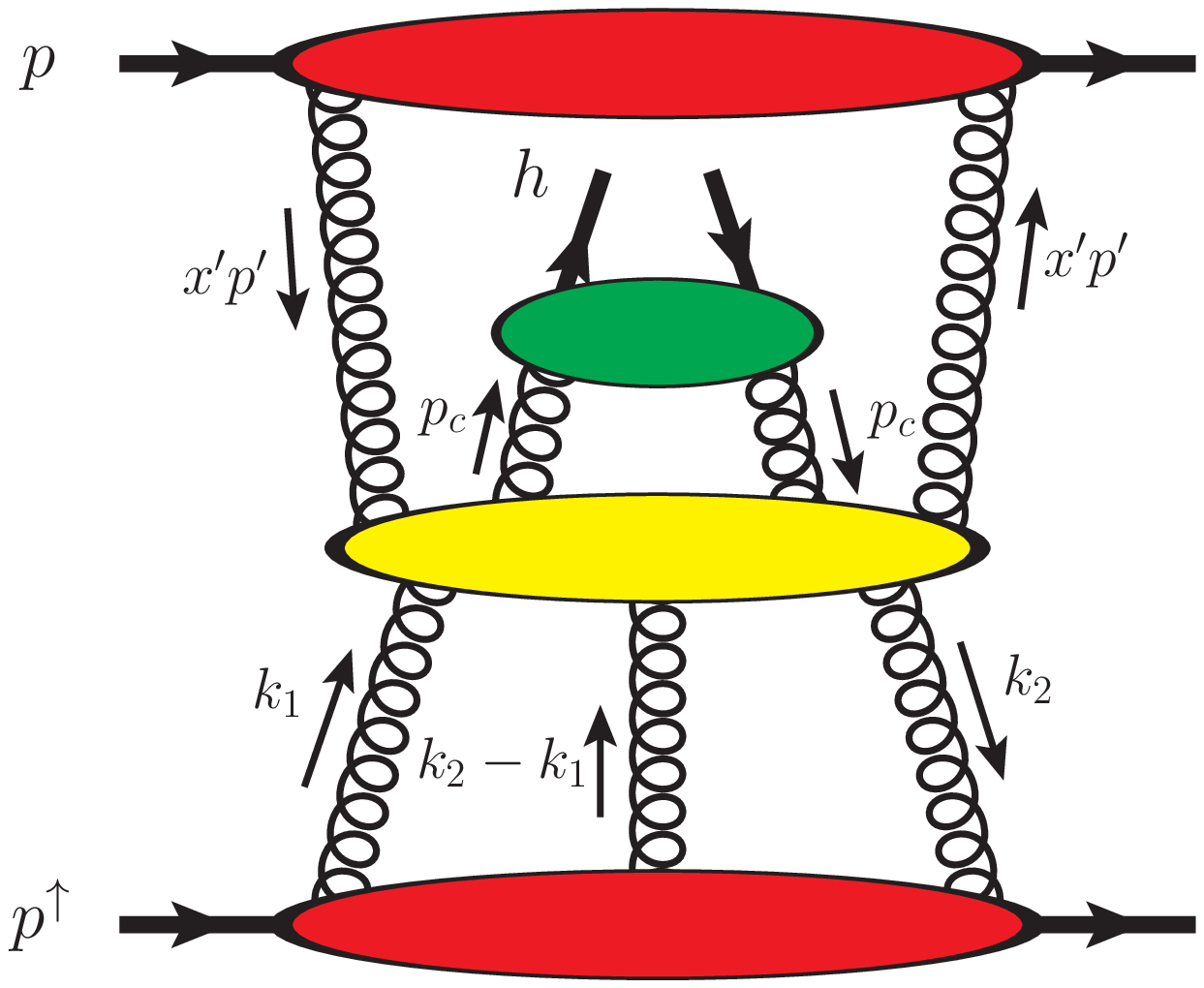}
\end{center}
 \caption{Schematic diagrams for the
contribution of the three-gluon correlation functions to 
the SSA in the light-hadron production $p^\uparrow p\to hX$.
The four blobs in each diagram represent, from the bottom to top,
three-gluon correlation (\ref{AAA}), the partonic hard part (\ref{hard1}),
the fragmentation function for the final hadron 
and the unpolarized parton distribution 
in the initial nucleon, respectively.}
\end{figure}

\begin{figure}[ht]
\begin{center}
  \includegraphics[height=8.5cm,width=15cm]{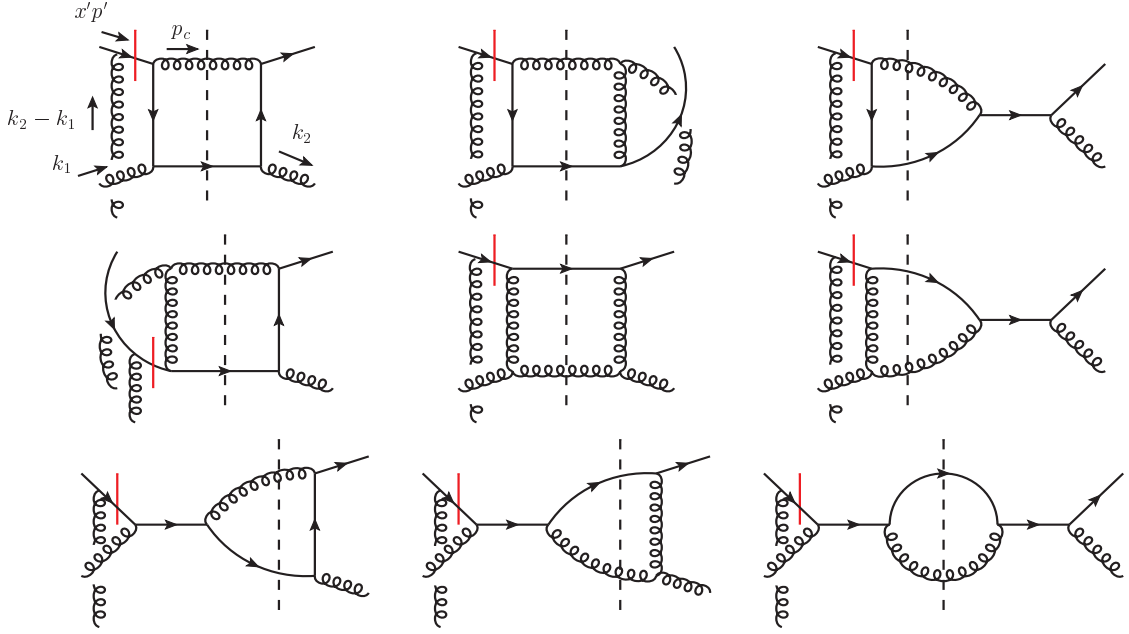}
\end{center}
 \caption{ISI diagrams in the gluon-fragmentation channel with the 
unpolarized quark distribution.
Shift of the momentum $p_c$ from the cut gluon-line to the cut quark-line
produces diagrams in
the quark-fragmentation channel with the 
unpolarized quark distribution, and the cross sections in these two channels are
connected by
$\hat{t}\leftrightarrow \hat{u}$.  }
\end{figure}

\begin{figure}[ht]
\begin{center}
  \includegraphics[height=8.5cm,width=15cm]{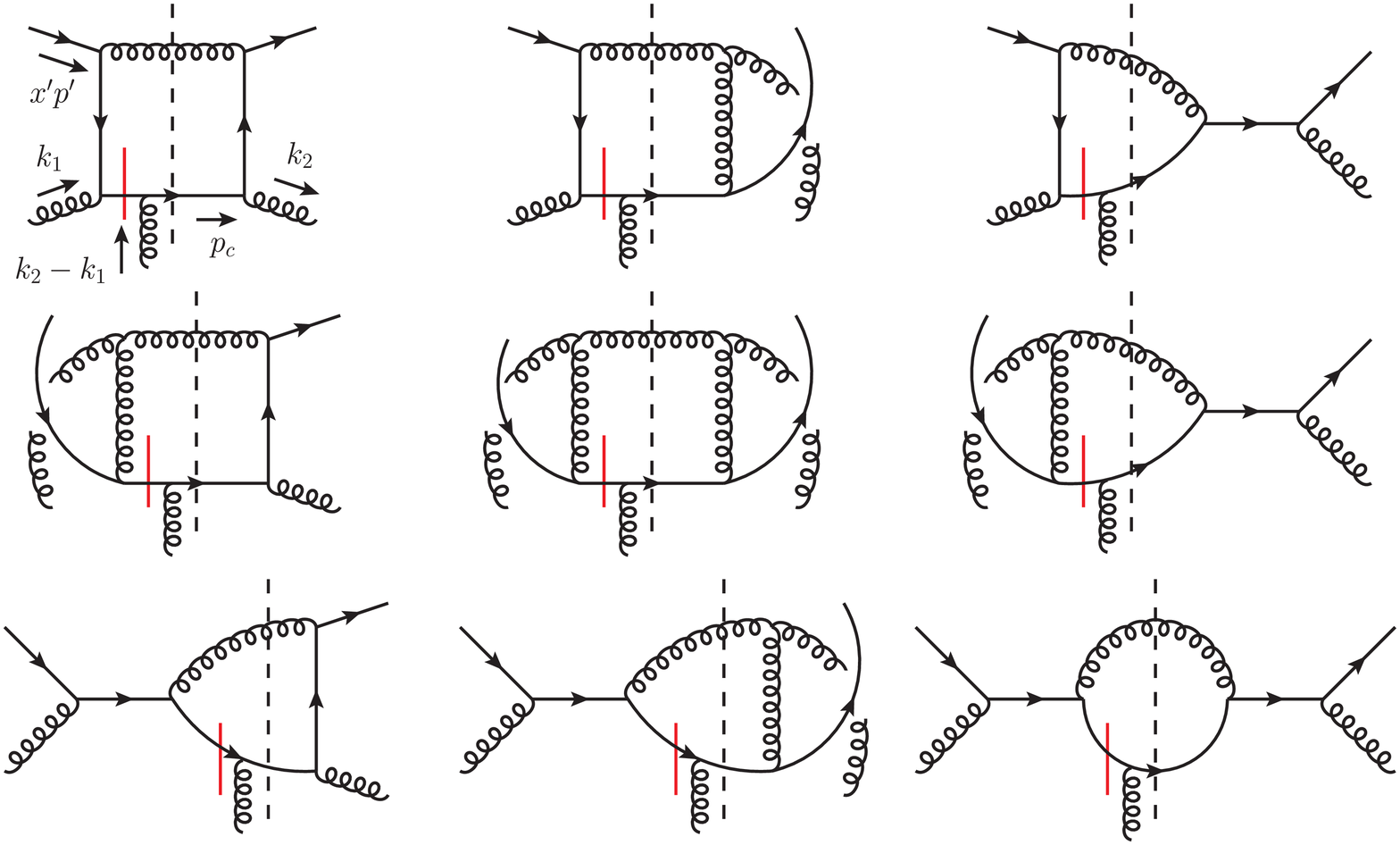}
\end{center}
 \caption{FSI diagrams in the quark-fragmentation channel with the 
unpolarized quark distribution.}
\end{figure}

\begin{figure}[ht]
\begin{center}
  \includegraphics[height=8.5cm,width=15cm]{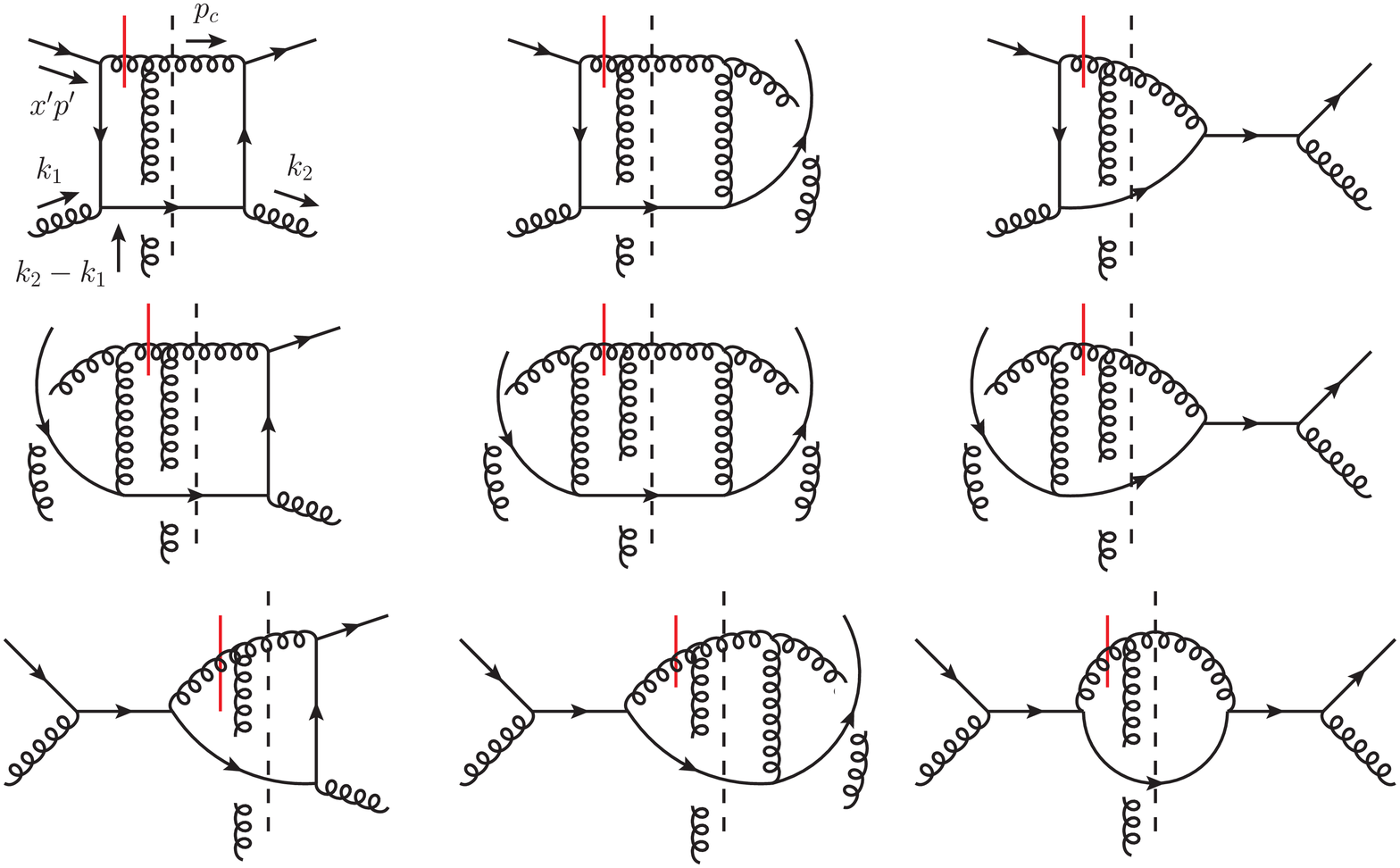}
\end{center}
 \caption{FSI diagrams in the gluon-fragmentation channel with the 
unpolarized quark distribution.}
\end{figure}

\begin{figure}[ht]
\begin{center}
  \includegraphics[height=8.5cm,width=15cm]{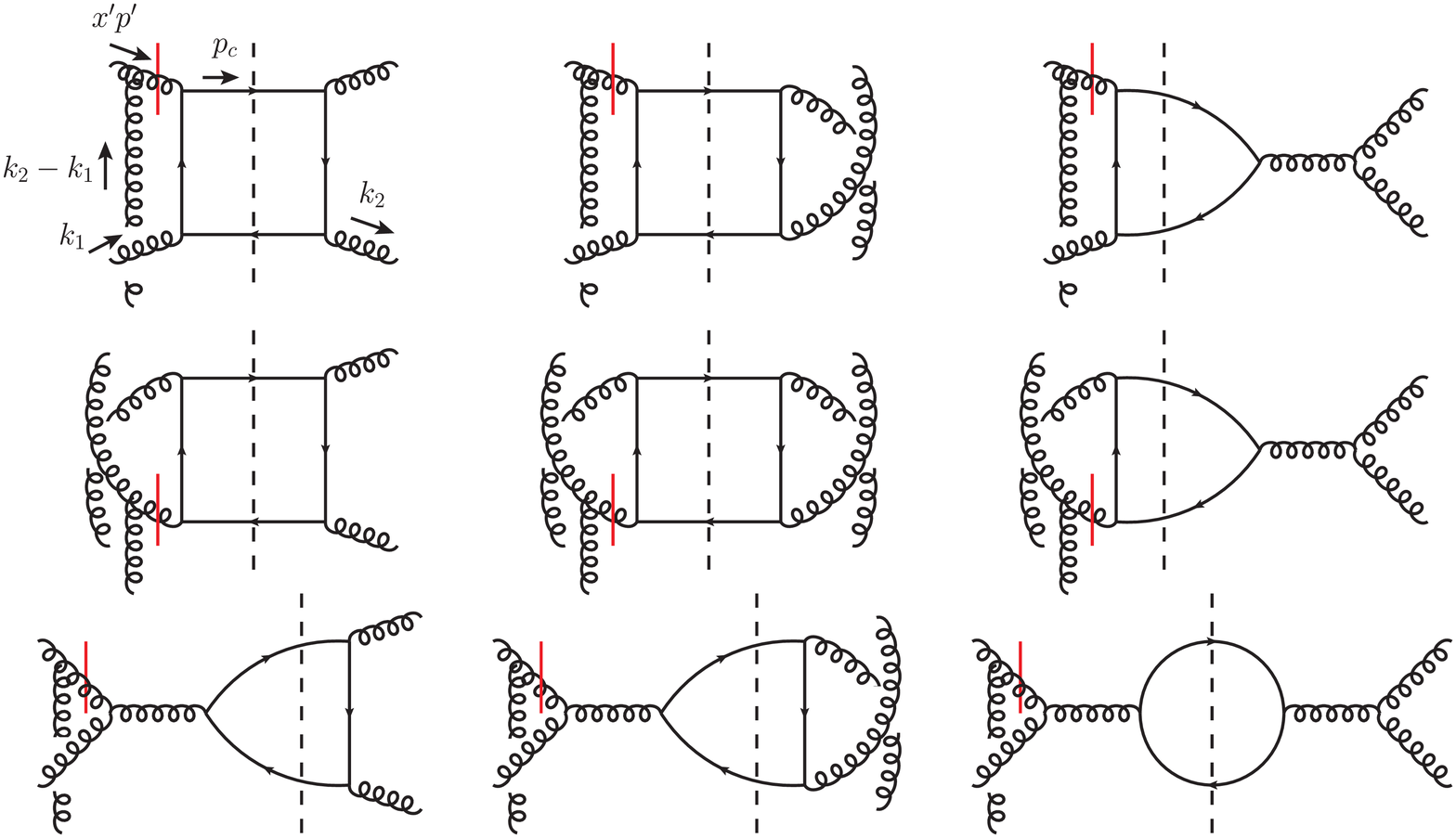}
\end{center}
 \caption{ISI diagrams in the quark-fragmentation channel with the 
unpolarized gluon distribution.}
\end{figure}

\begin{figure}[ht]
\begin{center}
  \includegraphics[height=8.5cm,width=15cm]{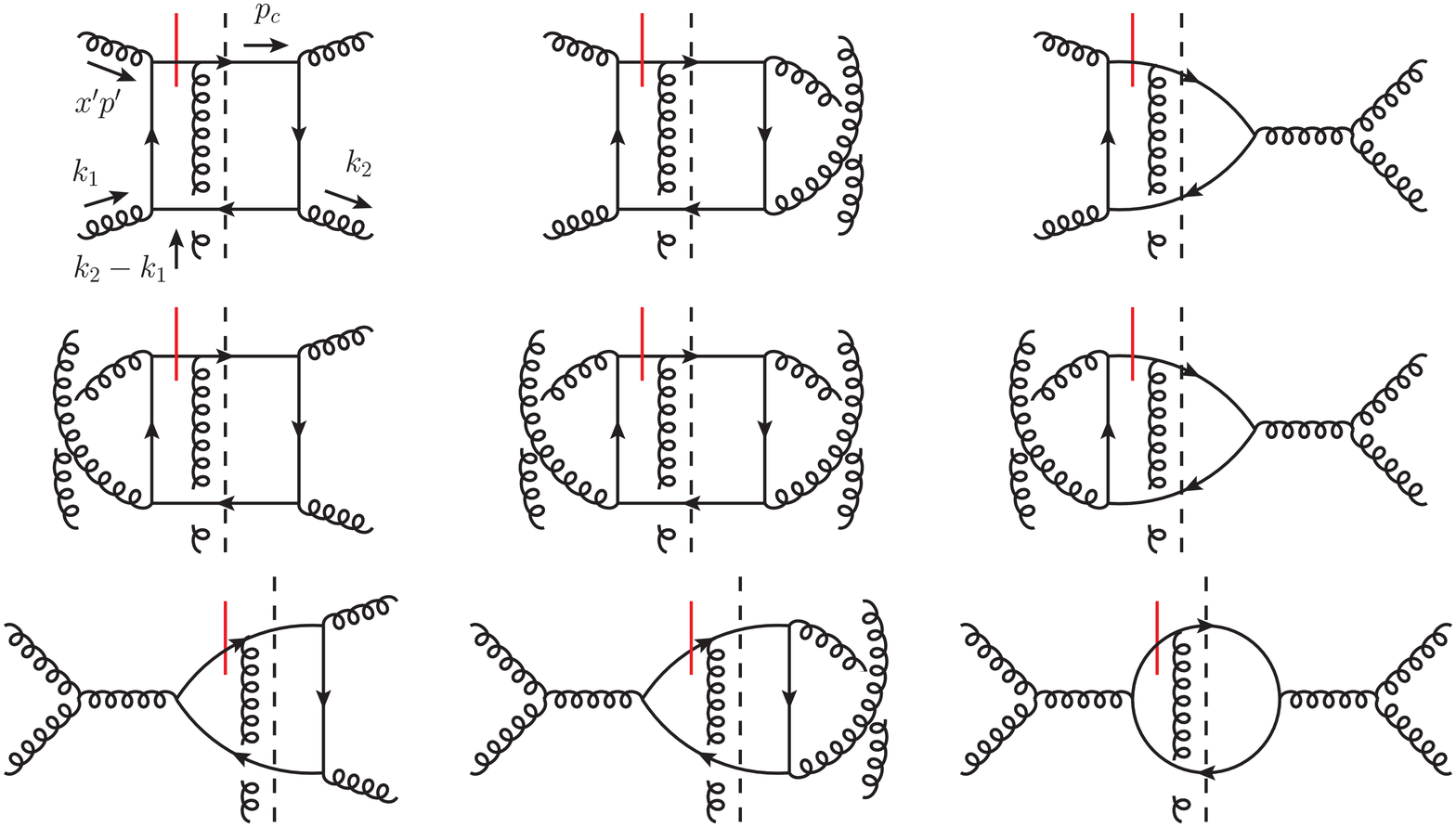}
\end{center}
 \caption{FSI diagrams in the quark-fragmentation channel with the 
unpolarized gluon distribution.}
\end{figure}

\begin{figure}[ht]
\begin{center}
  \includegraphics[height=8.5cm,width=15cm]{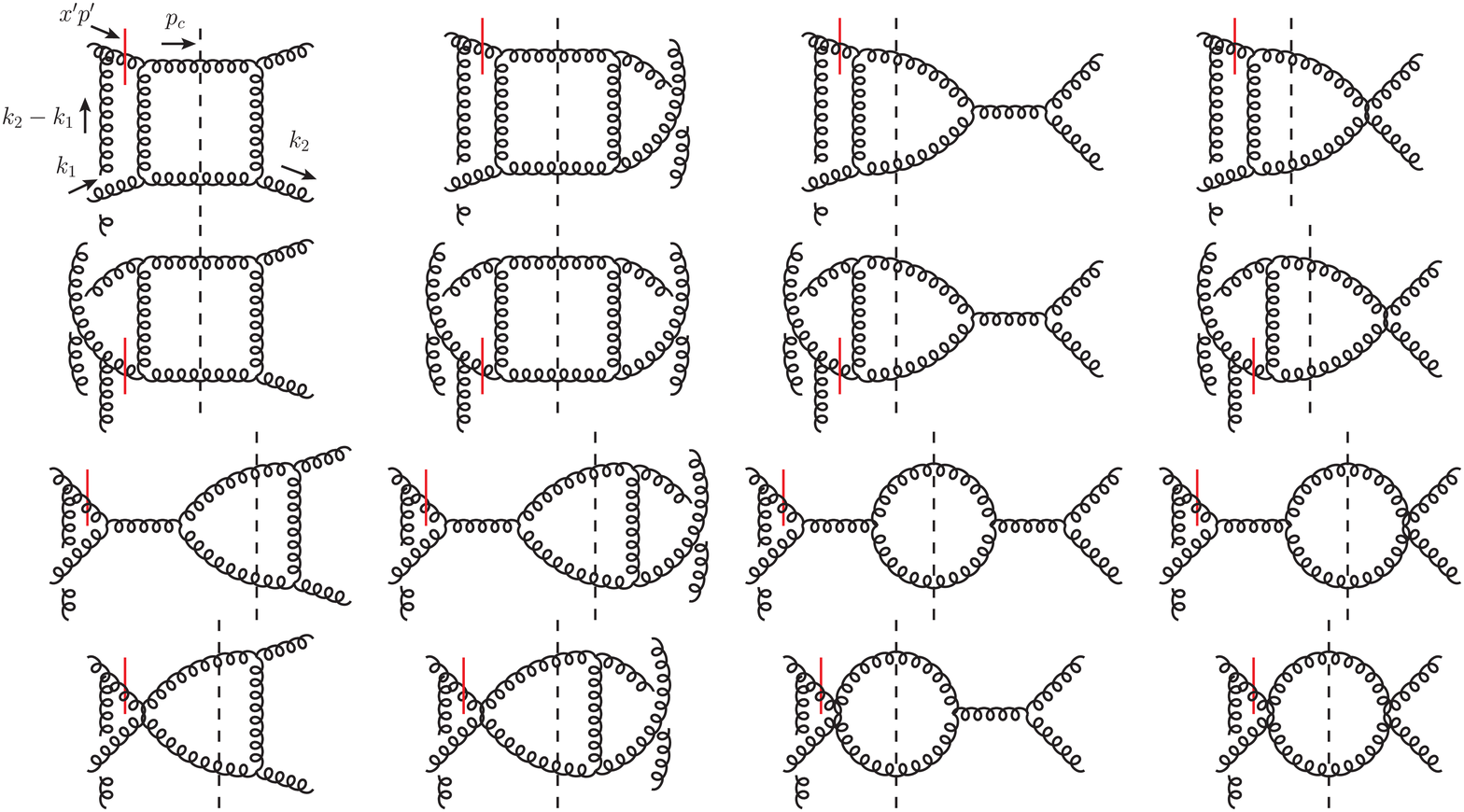}
\end{center}
 \caption{ISI diagrams in the gluon-fragmentation channel with the 
unpolarized gluon distribution.}
\end{figure}

\begin{figure}[ht]
\begin{center}
  \includegraphics[height=8.5cm,width=15cm]{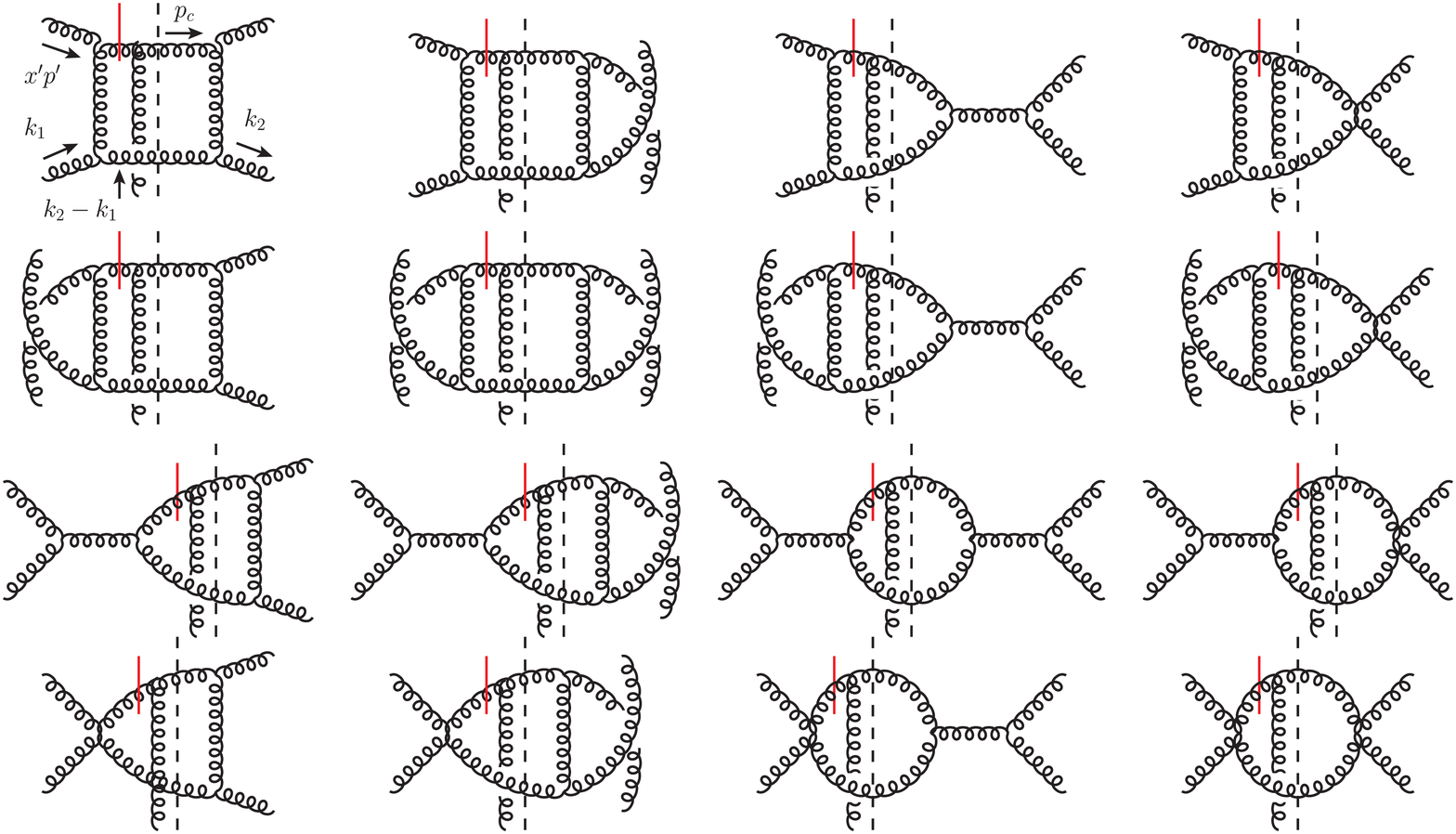}
\end{center}
 \caption{FSI diagrams in the gluon-fragmentation channel with the 
unpolarized gluon distribution.}
\end{figure}

The twist-3 single-spin-dependent cross section for
the process $p^\uparrow(p,S_\perp)+ p(p') \to h(P_h)+ X$
induced by the three-gluon correlation functions
can be obtained by applying the formalism developed 
in \cite{BeppuKoikeTanakaYoshida2010,KoikeTanakaYoshida2011,KoikeYoshida2011,KoikeYoshida2012}. 
Figure 1 shows the cut diagrams for the cross section.  
There the partonic hard part (second bottom blob),
\beq
S_{\mu\nu\lambda}^{abc}(k_1,k_2,x'p',p_c),
\label{hard1}
\eeq
the matrix elements in the polarized nucleon (bottom blob),
\beq
M_{abc}^{\mu\nu\lambda}(k_1,k_2)=g\int d^4\xi\int d^4\eta e^{ik_1\xi}e^{i(k_2-k_1)\eta}
\la pS_\perp | A_b^\nu(0) A_c^\lambda(\eta) A_a^\mu(\xi) |pS_\perp\ra,
\label{AAA}
\eeq
the unpolarized parton density (top blob) and the fragmentation function (second top blob)
are convoluted.   
In (\ref{hard1}) and (\ref{AAA}), $a$, $b$, $c$ are color indices, and $k_{1,2}$ 
are the momenta
of the gluon lines before collinear expansion as assigned in Fig. 1.
$x'p'$ is the momentum of the parton coming out of the unpolarized nucleon
and $p_c\equiv P_h/z$ is the one for the parton fragmenting into the final hadron.

Figures 2-8 represent the leading order (LO) diagrams for
$S_{\mu\nu\lambda}^{abc}(k_1,k_2,x'p',p_c)$.  
The real cross section occurs from a pole part
of the propagator with a short bar in the diagrams, 
which gives rise to the soft-gluon-pole 
(SGP) contributions at $x_1=x_2$.  
The diagrams have the structure that an extra gluon line with the 
momentum $k_2-k_1$ is attached
to either initial- or final- parton line in the diagrams 
representing the twist-2 cross section, 
and the propagator next to the attachment
gives a pole.  Therefore the diagrams are classified
as the initial-state-interaction (ISI) diagrams (Figs. 2, 5, 7) and the
final state interaction (FSI) diagrams (Figs. 3, 4, 6, 8).  
We shall employ the convention that  
the QCD coupling constant $g$ associated with the attachment of the extra gluon
to $S_{\mu\nu\lambda}^{abc}$ is included in the matrix element (\ref{AAA}) consistently with
the definition of the three-gluon correlation functions in (\ref{3gluonO}) and (\ref{3gluonN}).  
The mirror diagrams of Figs. 2-8 also contribute, and 
other pole contributions cancel among each other by taking the sum of the whole diagrams.

Applying the collinear expansion to the pole contribution of
$S_{\mu\nu\lambda}^{abc}(k_1,k_2,x'p',p_c)$, the LO
twist-3 cross section induced by the three-gluon correlation function is obtained as
\beq 
P^0_h\frac{d\Delta\sigma}{d^3P_h}&=&\frac{\alpha_s^2}{S}
\sum_{i,j}\int\frac{dx'}{x'}f_i(x')\int\frac{dz}{z^2}D_j(z)
\int\frac{dx_1}{x_1}\int\frac{dx_2}{x_2}\nonumber\\
&&\qquad\times
\left[
\left.
{\partial
S_{\mu\nu\lambda}^{abc}(k_1,k_2,x'p',p_c)p^{\lambda}\over 
\partial k_2^{\sigma}}\right|_{k_i=x_ip}
\right]^{\rm pole}
\omega^\mu_{\ \,\alpha}\omega^\nu_{\ \,\beta}\omega^\sigma_{\ \,\gamma}
M^{\alpha\beta\gamma}_{F,abc}(x_1,x_2), 
\label{twist3}
\eeq
where $\omega^\mu_{\ \,\alpha}=g^\mu_{\ \,\alpha}-p^\mu n_\alpha$, 
$f_i(x')$ and $D_i(z)$ are, respectively, unpolarized distribution
and fragmentation functions for the
quark and anti-quark flavors and the gluon ($i=q,\ \bar{q},\ g$), 
and 
$M^{\alpha\beta\gamma}_{F,abc}(x_1,x_2)$ is the
lightcone correlation function of the field-strengths defined as 
\beq
M^{\alpha\beta\gamma}_{F,abc}(x_1,x_2)
&=&-g(i)^3\int{d\lambda\over 2\pi}\int{d\mu\over 2\pi}e^{i\lambda x_1}
e^{i\mu(x_2-x_1)}\la pS_\perp|F_b^{\beta n}(0)F_c^{\gamma n}(\mu n)F_a^{\alpha n}(\lambda n)
|pS_\perp\ra \nonumber\\
&=&{N d_{bca}\over (N^2-4)(N^2-1)}
O^{\alpha\beta\gamma}(x_1,x_2)-{if_{bca}\over N(N^2-1)}N^{\alpha\beta\gamma}(x_1,x_2)
\label{Ffunction}
\eeq
with $O^{\alpha\beta\gamma}(x_1,x_2)$ and $N^{\alpha\beta\gamma}(x_1,x_2)$ 
defined in (\ref{3gluonO}) and (\ref{3gluonN}), respectively. 
The symbol $[\cdots]^{\rm pole}$ indicates the pole contribution is to be taken from 
the barred propagator in the hard part.  
In (\ref{twist3}), the factor $g^4$
is shifted to a prefactor from the hard part $S_{\mu\nu\lambda}^{abc}$.  
We remind that even though the analysis of Figs. 2-8 starts with
the gauge-noninvariant correlation function (\ref{AAA})
and the corresponding hard part $S_{\mu\nu\lambda}^{abc}(k_1,k_2,x'p',p_c)$, 
gauge-noninvariant contributions appearing in the collinear expansion
either vanish or cancel and 
the total surviving twist-3 contribution to the
single-spin-dependent cross section can be expressed as in (\ref{twist3}), using the
gauge-invariant correlation functions (\ref{3gluonO}) and (\ref{3gluonN}).

By calculating 
 $\left[\left.
{\partial
S_{\mu\nu\lambda}^{abc}(k_1,k_2,x'p',p_c)p^{\lambda}/
\partial k_2^{\sigma}}\right|_{k_i=x_ip}\right]^{\rm pole}$ from Figs. 2-8
contracted with
the coefficient tensors in the decomposition of (\ref{3gluonO}) and (\ref{3gluonN}), one obtains 
the twist-3 single-spin-dependent cross section as
\beq
E_{P_h}{d^3\Delta\sigma\over d^3P_h}&=&{2\pi M_N\alpha_s^2\over S}
\epsilon^{P_h p n S_{\perp}}
\sum_{i,j}\int {dx\over x}\int {dx'\over x'}f_i(x')\int {dz\over
z^2}D_j(z)\delta
(\hat{s}+\hat{t}+\hat{u}){1\over z\hat{u}}
 \nonumber\\
&&\times\biggl[\zeta_{ij}
\Bigl(\frac{d}{dx}O(x)-\frac{2O(x)}{x}
\Bigr)
\hat{\sigma}^{(O)}_{gi\to j} 
+\Bigl(\frac{d}{dx}N(x)-\frac{2N(x)}{x}\Bigr)
\hat{\sigma}^{(N)}_{gi\to j}
\biggr],   
\label{result}
\eeq
where the functions $O(x)$ and $N(x)$ are defined as
\beq
O(x)\equiv O(x,x) +O(x,0),\qquad N(x)\equiv N(x,x)-N(x,0), 
\label{OandN}
\eeq
and the factor $\zeta_{ij}$ is 
defined so that 
$\zeta_{ij}=-1$ when $i$ or $j$ is an anti-quark flavor
and $\zeta_{ij}=1$ for other cases.  
The partonic hard cross sections $\hat{\sigma}^{(O,N)}_{gi\to j}$
are the functions of the partonic 
Mandelstam variables defined as\footnote{The cross section (\ref{result}) is a function of
the Mandelstam variables $S=(p+p')^2$, $T=(p-P_h)^2$ and $U=(p'-P_h)^2$. }
\beq
\hat{s}=(xp+x'p')^2,\qquad \hat{t}=(xp-p_c)^2,\qquad \hat{u}=(x'p'-p_c)^2, 
\eeq
and 
can be written as the sum of the contributions from the ISI and FSI diagrams:
\beq
\hat{\sigma}^{(O,N)}_{gi\to j}=\hat{\sigma}^{(O,N)I}_{gi\to j}-{\hat{s}\over
\hat{t}}\hat{\sigma}^{(O,N)F}_{gi\to j}.  
\label{Iand F}
\eeq
They are given as follows:

\noindent
(i) Unpolarized quark distribution channels:\\

(a) Quark fragmentation channel (Figs. 2, 3):
\beq
\hat{\sigma}^{(O)I}_{gq\to q}=-\hat{\sigma}^{(N)I}_{gq\to
q}={\hat{t}(\hat{s}^2+\hat{t}^2)\over \hat{s}\hat{u}^2}-{1\over
N^2}\Bigl({\hat{s}\over \hat{t}}+{\hat{t}\over \hat{s}}\Bigr), 
\label{qfraqdis1}\\
\hat{\sigma}^{(O)F}_{gq\to q}=\hat{\sigma}^{(N)F}_{gq\to
q}=-{\hat{s}(\hat{s}^2+\hat{t}^2)\over \hat{t}\hat{u}^2}+{1\over
N^2}\Bigl({\hat{s}\over \hat{t}}+{\hat{t}\over \hat{s}}\Bigr).
\label{qfraqdis2}
\eeq

(a) Gluon fragmentation channel (Figs. 2, 4) :
\beq
&&\hat{\sigma}^{(O)I}_{gq\to g}=-\hat{\sigma}^{(N)I}_{gq\to g}
={\hat{u}(\hat{s}^2+\hat{u}^2)\over \hat{s}\hat{t}^2}-{1\over
N^2}\Bigl({\hat{s}\over \hat{u}}+{\hat{u}\over \hat{s}}\Bigr), 
\label{gfraqdis1}\\
&&\hat{\sigma}^{(O)F}_{gq\to
g}={(\hat{s}-\hat{u})(\hat{s}^2+\hat{u}^2)\over
\hat{s}\hat{t}\hat{u}}, 
\label{gfraqdis2}\\
&&\hat{\sigma}^{(N)F}_{gq\to
g}={(\hat{s}^2+\hat{u}^2)^2\over
\hat{s}\hat{t}^2\hat{u}}, 
\label{gfraqdis3}
\eeq

\newpage
\noindent
(ii) Unpolarized gluon distribution channels:\\

(b) Quark fragmentation channel (Figs. 5, 6):
\beq
&&\hat{\sigma}^{(O)I}_{gg\to q}={1\over C_F}{(\hat{u}-\hat{t})(\hat{t}^2+\hat{u}^2)
\over 2\hat{s}\hat{t}\hat{u}},
\label{qfragdis1}\\
&&\hat{\sigma}^{(N)I}_{gg\to q}={1\over C_F}{(\hat{t}^2+\hat{u}^2)^2
\over 2\hat{s}^2\hat{t}\hat{u}},
\label{qfragdis2}\\
&&\hat{\sigma}^{(O)F}_{gg\to q}=  \hat{\sigma}^{(N)F}_{gg\to q}=
\left( -{1\over C_F}{\hat{u}\over 2\hat{s}^2\hat{t}}
+{1\over N^2 C_F}{1\over 2\hat{t}\hat{u}}\right) (\hat{t}^2+\hat{u}^2). 
\label{qfragdis3}
\eeq

(b) Gluon fragmentation channel (Figs. 7, 8):
\beq
&&\hat{\sigma}^{(O)I,F}_{gg\to g}=0, 
\label{gfragdis1}\\
&&\hat{\sigma}^{(N)I}_{gg\to
g}={N\over C_F}{2(\hat{t}^2+\hat{u}^2)(\hat{t}^2+\hat{t}\hat{u}+\hat{u}^2)^2\over
\hat{s}^2\hat{t}^2\hat{u}^2}, 
\label{gfragdis2}\\
&&\hat{\sigma}^{(N)F}_{gg\to
g}=-{N\over C_F}{2(\hat{t}^2+2\hat{t}\hat{u}+2\hat{u}^2)(\hat{t}^2+\hat{t}\hat{u}+\hat{u}^2)^2\over
\hat{s}^2\hat{t}^2\hat{u}^2}. 
\label{gfragdis3}
\eeq
A remarkable feature of (\ref{result}) is that
the partonic hard cross sections for $O(x,x)$ and $O(x,0)$ are identical, 
and likewise for $N(x,x)$ and $-N(x,0)$.  Therefore
they contribute to the cross section through
the combinations $O(x)$ and $N(x)$ in (\ref{OandN}).   
The same feature was also observed for the twist-3 cross section for
the prompt-photon production $p^\uparrow p\to\gamma X$. 
Note that (\ref{qfragdis1}), (\ref{qfragdis2}) and (\ref{qfragdis3})
are obtained from the result for $p^\uparrow p\to DX$ in \cite{KoikeYoshida2011}
by taking the massless limit of the charm-quark mass, $m_c\to 0$.  
For the processes in which all partons participating in the 
scattering are massless, three-gluon correlations contribute
in the combination of $O(x)$ and $N(x)$.  
This is in contrast to the case for SIDIS and Drell-Yan processes,
where the virtual photon with large $Q^2$ enters the scattering.  
The result $\hat{\sigma}^{(O)I,F}_{gg\to g}=0$ in (\ref{gfragdis1})
is due the vanishing color factors.

\section{Numerical calculation of the asymmetry at the RHIC energy}

To illustrate the effect of the three-gluon correlation functions to $A_N$
for the light-hadron production,
we will present a model calculation of $A_N$ at the RHIC energy.  
To this end, 
we employ the same models which were used
for the study of
$A_N$ in $p^\uparrow p\to DX$ and $p^\uparrow p\to \gamma X$\,
\cite{KoikeYoshida2011,KoikeYoshida2012}.  
They are parametrized by using the twist-2 unpolarized gluon 
density $G(x)$ as  
\beq
&&{\rm Model\ 1}:\quad O(x)=N(x)=0.004\,x\,G(x),\label{model1}\\
&&{\rm Model\ 2}:\quad O(x)=N(x)=0.001\sqrt{x}\,G(x),\label{model2} 
\eeq
and the scale dependence of the three-gluon correlation is
also assumed to follow these
relations.\footnote{The evolution equation for the three-gluon correlation functions
$O(x_1,x_2)$ and $N(x_1,x_2)$
has been derived in \cite{Braun09}.  For this first rough estimate of $A_N$, however,
we use a simplified model as above.}
The coefficients 0.004 and 0.001 
were determined so that
the calculated $A_N^D$ does not exceed the RHIC preliminary data for 
$A_N^D$\,\cite{Liu}.  
The above model ansatz was motivated to see the effect of the three-gluon correlations
in comparison with the gluon density and to see the sensitivity of $A_N$
to the small-$x$ behavior of the functions.  
We use the unpolarized parton density in \cite{GJR08} 
and the fragmentation function for the pion in \cite{FlorianSassotStratmann2007PIK}.  
For the calculation, we set the scale of all the distribution and fragmentation functions
at the transverse momentum of the final hadron $P_T$.

\begin{figure}[ht]
\begin{center}
  \includegraphics[height=6cm,width=8cm]{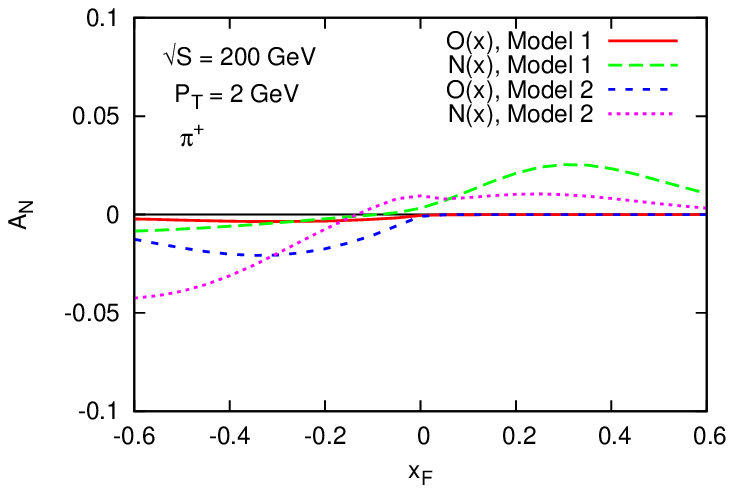}
  \includegraphics[height=6cm,width=8cm]{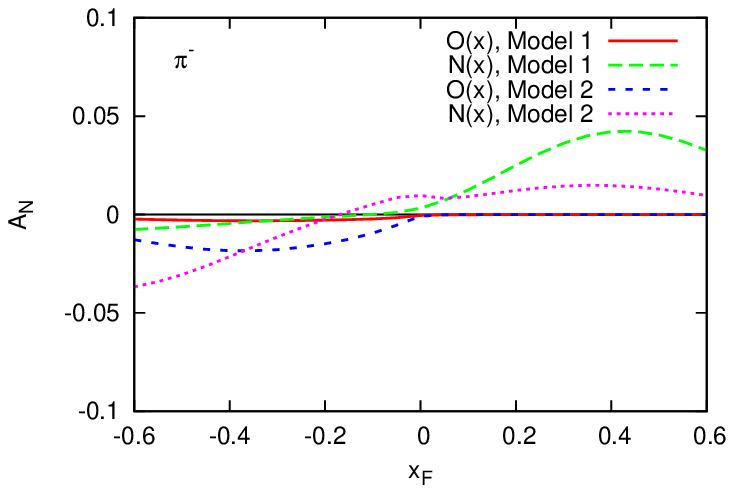}
  \includegraphics[height=6cm,width=8cm]{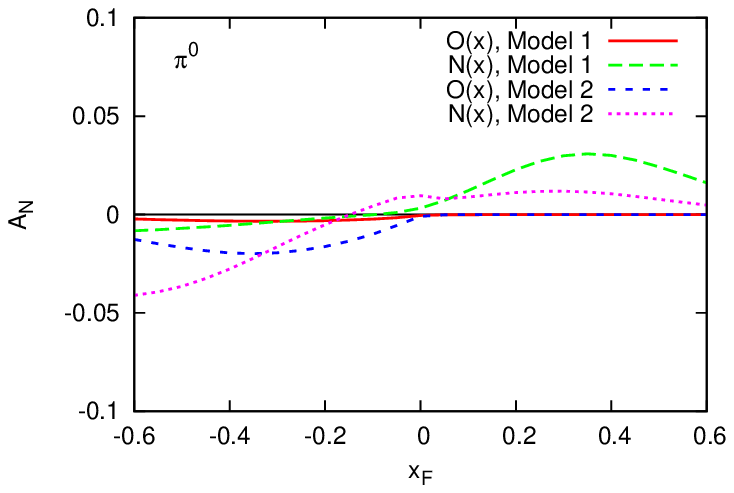}
  \includegraphics[height=6cm,width=8cm]{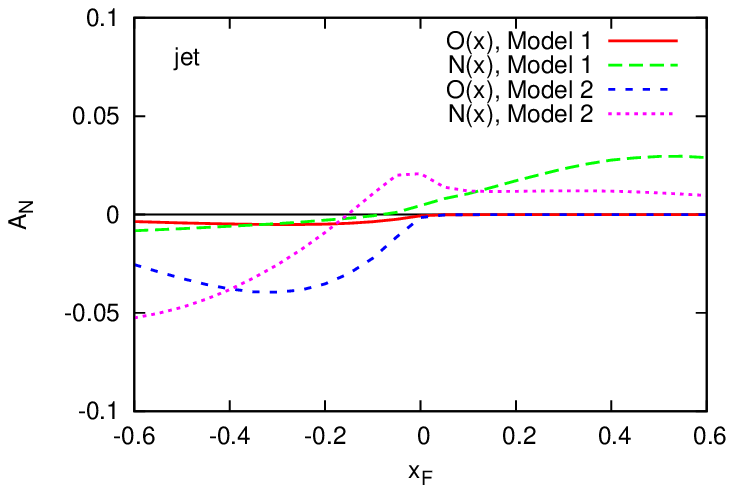}
\end{center}
 \caption{$x_F$-dependence of the 
three-gluon contribution to $A_N$ for $p^\uparrow p\to \{\pi^{\pm,0},{\rm jet}\}X$
at $\sqrt{S}=200$ GeV
and $P_T=2$ GeV.  The contribution from $O(x)$ and $N(x)$ are plotted separately
for models 1 and 2.}
\end{figure}

\begin{figure}[ht]
\begin{center}
  \includegraphics[height=6cm,width=8cm]{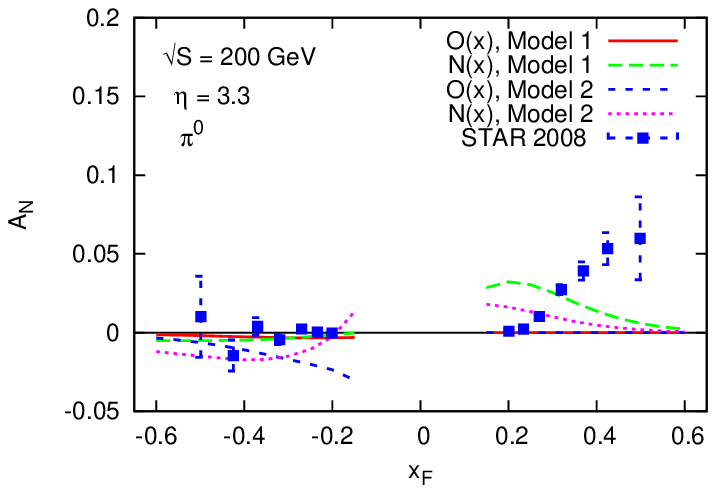}
  \includegraphics[height=6cm,width=8cm]{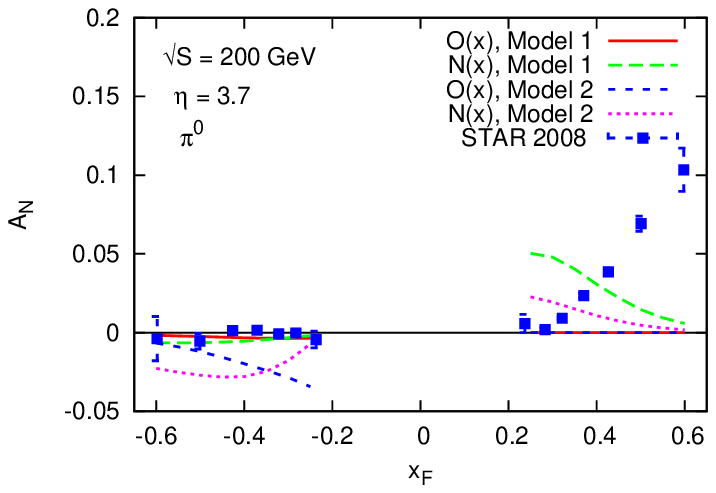}
\end{center}
 \caption{$x_F$-dependence of the three-gluon 
contribution to $A_N$ for $p^\uparrow p\to \pi^{0} X$
at $\sqrt{S}=200$ GeV
and $\eta=3.3$ and 3.7
in comparison to the RHIC-STAR data\,\cite{Star2008}.
The contribution from $O(x)$ and $N(x)$ are plotted separately
for models 1 and 2.}
\end{figure}

Figure 9 shows the $x_F$-dependence of the three-gluon contribution 
to $A_N$ for the $\pi^{\pm,0}$ and jet productions in the pp-collision
at the RHIC energy $\sqrt{S}=200$ GeV and 
$P_T=2$ GeV.   We plotted the contribution from
$O(x)$ and $N(x)$ separately to see each effect on $A_N$.  
At $x_F>0$, $N(x)$ gives rise to the large asymmetry 
and the effect of $O(x)$ is negligible for both models.
The origin of the large asymmetry from $N(x)$ is
the 
partonic cross section (\ref{gfragdis3}): 
At large $x_F>0$, where $-T\ll S\sim -U$, 
large-$x$ and small-$x'$ region 
is probed, and thus the unpolarized gluon density 
brings main contribution.  
While 
the partonic cross sections in the quark fragmentation channel
(\ref{qfragdis1})-(\ref{qfragdis3}) are tiny, 
those in the gluon fragmentation channel 
(\ref{gfragdis2}) and (\ref{gfragdis3}) for $N(x)$
are large, in particular, the latter contribution is
enhanced by the kinematic factor $\hat{s}/\hat{t}$ in (\ref{Iand F})
for the FSI.

At $x_F<0$, in particular, $x_F\to -1$ where $-U\ll S\sim -T$, 
the region of small-$x$ and large-$x'$ is relevant. 
Thus the model 1 gives rise to only small asymmetry for
both $O(x)$ and $N(x)$ due to their mild behavior at small-$x$.  
On the other hand, 
the model 2 gives the large asymmetry
for the two functions.  
This is
due to the large partonic cross section (\ref{qfraqdis1}) and (\ref{qfraqdis2})
in the unpolarized quark distribution channel with the quark fragmentation, 
and a steeply rising behavior in
the three-gluon correlations
at small-$x$, which is even more enhanced by the
derivative.   
For $O(x)$, there is a partial cancelation between the ISI and FSI due to the relative signs 
between (\ref{qfraqdis1}) and (\ref{qfraqdis2}),
while for $N(x)$ these cross sections contribute constructively.
This leads to different behavior of the asymmetry between $N(x)$ and $O(x)$.

Figure 10 shows the three-gluon contribution to $A_N^{\pi^0}$
in comparison to the RHIC-STAR data\,\cite{Star2008} at $\sqrt{S}=200$ GeV and the 
pseudorapidity $\eta=3.3$ and 3.7. 
One sees that the contribution from $N(x)$
is much larger than the data at small $x_F>0$ for the two models, and therefore
it is unlikely that the magnitude of $N(x)$ is as large as 
these models in the large-$x$ region.  
At $x_F<0$, the contribution of the model 1 is zero and is consistent with data
for both $N(x)$ and $O(x)$, while the model 2 for the two functions is far from the data points.
This means the three-gluon correlations should behave more mildly than the model 2
in the small-$x$ region.  
We note, however, that
the observed asymmetry results from the combination of the
quark-gluon correlation function, twist-3 fragmentation function
and the three-gluon correlation function,
and thus we need a complete analysis including all these effects
to draw a definite conclusion.  

Figure 11 shows the calculated $A_N^{\pi^0}$ in the
midrapidity region ($|\eta|<0.35$) at $\sqrt{S}=200$ GeV in comparison with
the RHIC-PHENIX data\,\cite{Phenix2013}.  Both models give tiny asymmetry due to the small
partonic cross sections, so the form of the three-gluon correlation functions
is not much constrained by the data in this region.

\begin{figure}[ht]
\begin{center}
  \includegraphics[height=7cm,width=10cm]{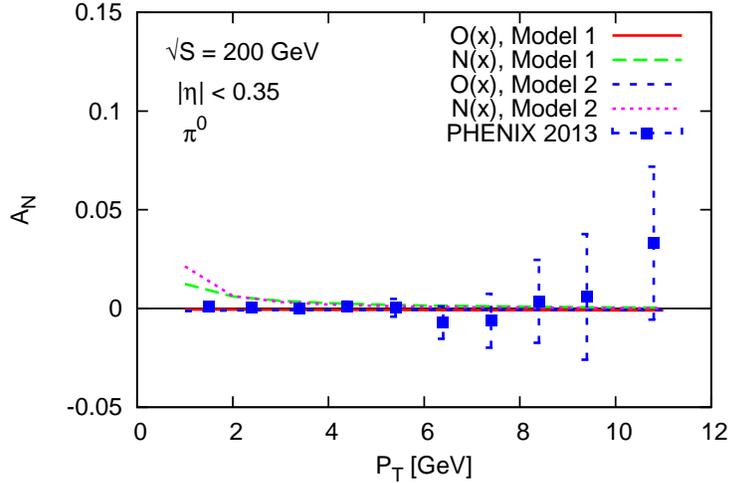}
\end{center}
 \caption{$P_T$-dependence of the three-gluon 
contribution to $A_N$ for $p^\uparrow p\to \pi^{0} X$
at $\sqrt{S}=200$ GeV and $|\eta|<0.35$
in comparison to the RHIC-PHENIX data\,\cite{Phenix2013}.
The contribution from $O(x)$ and $N(x)$ are plotted separately
for models 1 and 2.}
\end{figure}

\section{Summary}

In this paper we have studied the three-gluon contribution to SSA for the
light hadron production in the pp collision, $p^\uparrow p\to hX$.  
We have derived the corresponding LO twist-3 cross section.  
Together with the result for the contribution from
the quark-gluon correlation and the twist-3 fragmentation functions,
this has completed the twist-3 cross section for this process.  
We have also presented a numerical calculation of the asymmetry
at the RHIC energy based on our previous models 
and have shown that this process could bring a useful constraint
on the upper bound of the three-gluon correlation functions.  

\section*{Acknowledgments}

The work of K.K. is supported by the Grand-in-Aid for
Scientific Research (No.24.6959) from the Japan Society
of Promotion of Science.
The work of Y.K. is supported in part by the Grant-in-Aid for
Scientific Research
(No.23540292) from the Japan Society of Promotion of Science.
The work of S.Y. is supported by
JSPS Strategic Young Researcher Overseas Visits Program
for Accelerating Brain Circulation (No.R2411)
.

\end{document}